\documentclass[amsmath,amssymb,aps,twocolumn]{revtex4-2}

\usepackage{graphicx}
\usepackage{bm}
\usepackage{url}
\usepackage{color}
\usepackage{hyperref}
\hypersetup{
	pdfauthor={Daniel Jampolski, Luciano Rezzolla},
	pdftitle={Formation of gravastars},
	colorlinks=true,
	linkcolor=blue,
	citecolor=cyan,
}
\usepackage{mathtools}
\usepackage{times}
\usepackage{mathrsfs}
\usepackage{enumitem}

\newcommand{\ie}{i.e.,~}
\newcommand{\eg}{e.g.,~}

\newcommand{\defeq}{\vcentcolon=}
\newcommand{\eqdef}{=\vcentcolon}

\newcommand\ShiftDown[2]{\raisebox{-#1}{\upshape #2}}

\begin{document}
\title{Formation of gravastars}

\author{Daniel Jampolski\vspace{0.5em}}
\affiliation{Institut f{\"u}r Theoretische Physik,
  Max-von-Laue-Strasse\,{1}, 60438 Frankfurt, Germany}

\author{Luciano Rezzolla\vspace{0.5em}}
\affiliation{Institut f{\"u}r
  Theoretische Physik, Max-von-Laue-Strasse\,{1},60438 Frankfurt,
  Germany\vspace{0.25em}}
\affiliation{School of Mathematics, Trinity College, Dublin 2,
  Ireland\vspace{0.25em}}
\affiliation{CERN, Theoretical Physics Department, 1211 Geneva 23,
  Switzerland\vspace{1.0em}}

\begin{abstract}
Regular black holes and horizonless black hole mimickers offer
mathematically consistent alternatives to address the challenges posed by
standard black holes. However, the formation mechanism of these
alternative objects is still largely unclear and constitutes a
significant open problem since understanding their dynamical formation
represents a first step to assess their existence. We here investigate,
for the first time and without invoking higher-curvature corrections, the
dynamical formation of a well-known horizonless black hole mimicker,
namely, a gravastar. More specifically, starting from the collapse of a
uniform dust sphere as in the case of the Oppenheimer-Snyder collapse, we
demonstrate that, under fine-tuned conditions, a gravastar can form from
the nucleation and expansion of a de Sitter region with initial zero size
at the center of the collapsing sphere. Furthermore, the de Sitter
expansion naturally slows down near the Schwarzschild radius, where it
meets the collapsing dust surface and gives rise to a static
equilibrium. Interestingly, we also find a maximum initial compactness of
the collapsing star of $\mathcal{C}= 3/8$, above which the collapse to a
black hole is inevitable.
\end{abstract}

\maketitle

\noindent\textit{Introduction}. Black holes are important cornerstones
of modern astrophysics and the end point of gravitational
collapse~\cite{Penrose65, Penrose70a}. While their existence is perfectly
compatible with gravitational-wave~\cite{Abbott2016fw} and
electromagnetic~\cite{Akiyama2019_L1_etal, EHT_SgrA_PaperI_etal}
observations, some of the properties of black holes remain a source of
concern and debate. Classical black holes, in fact, contain physical
singularities where the predictability of general relativity breaks
down. Additionally, they are shielded by an event horizon whose
semiclassical evolution leads to problems with loss of information. To
resolve these issues, alternative mathematical descriptions of black
holes have gained significant attention over the last decades. These
alternatives either remove the singularity while preserving the existence
of an event horizon---as in the case of so-called ``regular black
holes''---or do not have either---as in the case of ``black hole
mimickers'' (see, \eg \cite{Bardeen68, Dymnikova1992, Hayward2006PRL,
  Simpson2019, Gambini2020, Lobo2021, Ashtekar2023, Maso2023,
  Bonanno2024, Bueno2024, Casadio2024, Khodabakhshi2025, Casadio2025,
  Harada2025} for some regular black hole solutions or
\cite{Carballo2025} for a recent review).

It is in this general context that the ingenious model of the
gravitational vacuum condensate star, or gravastar, was first
proposed~\cite{Mazur2001published, Mottola2023}. It comprises an interior
of dark energy, described by the de Sitter solution and stabilized by a
thin shell of ordinary matter. While the negative pressure of the dark
energy seeks to expand the interior, gravity and a tangential pressure in
the shell act as a belt preventing the expansion and leading to a static
solution. Because a gravastar possesses neither a singularity nor an
event horizon, and since its compactness can be brought arbitrarily close
to that of a black hole, it has long been argued that it would be
difficult to distinguish it from a black hole. When considering
electromagnetic radiation as a route to distinguish the two objects, the
debate is still ongoing (see, \eg \cite{Broderick2007,
  EHT_SgrA_PaperVI_etal, Carballo2023b}), while it is clear that the
different response to gravitational perturbations makes gravastars
distinguishable from black holes~\cite{Chirenti2007, Chirenti2016}. An
aspect of gravastars that has so far not been addressed, mostly because
of the challenges it poses, is their genesis from generic spherical
distribution of matter (see however~\cite{Nakao2018} for gravastar
formation from the collapse of matter shells). We here present, for the
first time, a model for the creation of a static gravastar following a
gravitational collapse of a spherical cloud of matter.

\vspace{\baselineskip}
\noindent\textit{Arrested Oppenheimer-Snyder Collapse}. Our starting
point for the construction of a gravastar formation model is the
Oppenheimer-Snyder (OS) collapse~\cite{Oppenheimer39a}. This well-known
solution, which physically describes the collapse of a homogeneous dust
sphere to a black hole, is mathematically modeled by a
Friedmann-Lemaitre-Robertson-Walker (FLRW) solution matched to an
exterior Schwarzschild solution. In our case, since we eventually want to
obtain a static gravastar, we need to add an expanding de Sitter region
inside the collapsing FLRW spacetime, noting that the appearance of a
de Sitter solution represents a common feature of most ``bouncing''
solutions in gravitational collapse (see, \eg \cite{Lobo2021,
  Ashtekar2023, Maso2023, Bueno2024}).

Thus, at any time in the evolution we separate spacetime into three
different regions, I--III, with the line element for each region being
given by
\begin{align}
  &d s_{\rm I}^2 = - d T^2+a^2_{\rm I}(T)\left[{ d R^2}/\left({1-k_{\rm
        I}R^2}\right)+R^2 d \Omega^2\right]\,,\\
  &d s_{\rm II}^2 = - d \tau^2+a^2_{\rm II}(\tau)\left[{ d
      \rho^2}/\left({1-k_{\rm II}\rho^2}\right)+\rho^2 d
    \Omega^2\right]\,,\\
  &d s_{\rm III}^2 = -\Phi(r) d t^2 + { d r^2}/{\Phi(r)}+r^2 d
  \Omega^2\,.
\end{align}
Each region has its own set of coordinates, so that region I is the
expanding FLRW de Sitter solution with $(T,R)$ as a set of comoving
coordinates for $0\leq R<R_1$, while region II is the contracting FLRW
dust solution with $(\tau,\rho)$ as a set of comoving coordinates for
$\rho_1(\tau)<\rho<\rho_2$, and, finally, region III is a vacuum
Schwarzschild solution with $(t,r)$ as a set of Schwarzschild coordinates
for $r>r_2(t)$.

Because the first two regions are described by FLRW solutions, they are
characterized by two distinct scale factors $a_{\rm I}$ and $a_{\rm II}$,
and by two spatial curvature constants $k_{\rm I}<0$ (expansion from zero
initial scale factor) and $k_{\rm II}>0$ (collapse from initial static
dust cloud), respectively. In the third region, instead, we have $\Phi(r)
\defeq 1-2M/r$ with $M$ being the Schwarzschild mass, which is a constant
of the spacetime. Without loss of generality, and because of the
underlying spherical symmetry, all regions share the same spherical
coordinates $(\theta, \phi)$ with $d \Omega^2 \defeq d \theta^{2} + {\rm
  sin}^2\theta\, d\phi^{2}$. Clearly, $R$ and $\rho$ are not areal
coordinates, while $r$ is.

A few comments should be made before proceeding further. First, due to
their comoving nature, $\rho_1=\rho_1(\tau)$ and $r_2=r_2(t)$, but also
that $R_1={\rm const.}$ and $\rho_2={\rm const.}$ since they are the
initial coordinate radii of the de Sitter bubble and of the dust sphere,
respectively. Second, we assume that the energy density $e$ and pressure
$p$ in the two FLRW spacetimes follow an equation of state $p_{\rm
  I}=-e_{\rm I}$ for the de Sitter region and $p_{\rm II}=0$ for the
collapsing dust region. Third, using the Friedmann equations it is
possible to obtain that $e_{\rm I}={\rm const.}$ and $e_{\rm
  II}=\bar{e}_{\rm II}(\bar{a}_{\rm II}/a_{\rm II})^3$, where
$\bar{e}_{\rm II}\defeq e_{\rm II}(0)$ and $\bar{a}_{\rm II} \defeq
{a}_{\rm II} (0)$. Fourth, the expressions for the scale factors can be
derived by solving the first of the Friedmann equations, obtaining
$a_{\rm I}(T)= \sqrt{3|k_{\rm I}|/{8\pi e_{\rm I}}} \,\,
\mathrm{sinh}\left(\sqrt{{8\pi e_{\rm I}}/{3}} \,\, T\right)$ for the
de Sitter scale factor, and $a_{\rm II}(\eta)={\bar{a}_{\rm
    II}}/{2}\,\left[1+{\rm cos}\left(\sqrt{k_{\rm
      II}}\,\eta\right)\right]$ for the dust scale factor, where we have
used the cycloid coordinate $\eta$ with a coordinate transformation $d
\eta=d\tau/a_{\rm II}$, so that $\tau(\eta) = {\bar{a}_{\rm
    II}}/{2}\,\left[\eta + \sqrt{1/k_{\rm II}}\,{\rm
    sin}\left(\sqrt{k_{\rm II}}\,\eta\right)\right]$. Fifth, it is useful
to introduce the proper circumferential radii of the hypersurfaces
dividing the three spacetimes, \ie $\mathcal{R}_1 \defeq {\mathscr{C}_1}
/ {2\pi} = a_{\rm I}(T)R_1$ and $\mathcal{R}_2 \defeq {\mathscr{C}_2} /
{2\pi} = a_{\rm II}(\eta) \rho_2$, where $\mathscr{C}_{1,2}$ are the
proper (polar) circumferences of the hypersurfaces. Finally, we assume
$a_{\rm I}(0)=0$ and $\partial_{\eta}a_{\rm II}(0)=0$, that is, the
de Sitter bubble is initially of zero size and the dust collapses with
zero initial velocity. We should note that while not uncommon in general
relativity (e.g., point particles) zero-size objects represent the
limitations of a classical theory. We here adopt it because it provides a
mathematical simplification, but the initial zero-size de Sitter bubble
could be replaced by a finite-size one without any impact on the dynamics
we will describe. Obviously, if a quantum-gravitational description were
possible, the zero-size de Sitter bubble would be naturally replaced by a
Planck-size bubble.

\vspace{\baselineskip}
\noindent\textit{Junction Conditions}. Since we are patching together
different solutions of the Einstein equations at the hypersurfaces
between each region, we need to be careful in handling these
separatrices. More specifically, we need to enforce junction conditions
ensuring that the metric induced on the hypersurfaces is continuous
across it~\cite{Israel1966}. Furthermore, it may be desirable that also
the extrinsic curvature is continuous, although this is not a necessary
condition for the patching to be valid. As in the OS collapse, the
extrinsic curvature in our model is continuous at the dust hypersurface
separating regions II and III. However, it is not continuous at the
de Sitter hypersurface separating regions I and II as a result of the
(infinite) radial-pressure gradient appearing when going from the
expanding de Sitter bubble to the collapsing dust region. The loss of
continuity is reflected by the appearance of a surface tension and of a
surface energy density~\cite{Rezzolla1994} on the de Sitter hypersurface.

While the junction conditions for the dust hypersurface can be found in
textbooks (see, \eg \cite{Poisson04a, Rezzolla_book:2013}), which fix
$\bar{e}_{\rm II}$, $r_2(t)$ and $t(\eta)$, the junction conditions for
the de Sitter hypersurface need to be worked out. Imposing the continuity
of the induced metric yields
\begin{equation}
  \rho_1(T)=R_1{a_{\rm I}(T)}/{a_{\rm II}\left(\eta(T)\right)}\,,
  \label{dS-position}
\end{equation}
and this gives the position of the de Sitter hypersurface as seen from an
observer in the dust region. Similarly, the relation between the
coordinate time in the collapsing dust region $\eta$ and that in the
expanding de Sitter region $T$ is given by
\begin{equation}
  \dfrac{d \eta}{d T}=\dfrac{R_{1}\rho_1 \dot{a}_{\rm I} a^{\prime}_{\rm
      II} \pm a_{\rm II} \sqrt{(k_{\rm II} \rho_1^2 -
      1)\left(\rho_1^2/\rho_{\star}^{2} - 1 - R_{1}^{2}\dot{a}^2_{\rm
        I}\right)}}{\ShiftDown{0.25em}{$ a_{\rm
        II}^{2}(\rho_1^2/\rho_{\star}^{2} - 1)$}}\,,
  \label{dtau-dT}
\end{equation}
where the choice of the minus sign corresponds to the positive time
direction in $\eta$ and where $\dot{a}_{\rm I} \defeq \partial_T a_{\rm
  I}$ and $a^{\prime}_{\rm II} \defeq \partial_{\eta} a_{\rm
  II}$. Satisfying Eqs.~\eqref{dS-position}--\eqref{dtau-dT}, together
with the junction conditions for the dust hypersurface, guarantees a
valid patching of the different regions of the spacetime and, of course,
that observers on either side on an hypersurface will measure the same
values of $\mathcal{R}_1$ and $\mathcal{R}_2$.

An interesting feature of Eq.~\eqref{dtau-dT} is the divergence in the
rate between the two times, which occurs when the de Sitter hypersurface
reaches the so-called ``critical radius'' $\rho_{\star}(\eta)$ defined as
\begin{equation}
  \rho_1(\eta)=\left[{{k_{\rm II} + \left(\dfrac{a^{\prime}_{\rm
            II}}{a_{\rm II}}\right)^{2}}}\right]^{-1/2} =
  \rho_2\sqrt{\dfrac{r_2}{2M}}\eqdef\rho_{\star}(\eta)\,.
  \label{critical-radius}
\end{equation}
When $\rho_1 \to \rho_{\star}$, the expansion of the de Sitter bubble
stops, as seen from an observer in the dust region, that is
\begin{equation}
  \frac{ d \eta}{ d T}\to\infty\,\,\,
  \Rightarrow\,\,\,\partial_{\eta}\,a_{\rm I}\left(T(\eta)\right) =
  \partial_{T}\,a_{\rm I}\left(\frac{d \eta}{dT}\right)^{-1}\to 0\,.
\end{equation}

\vspace{\baselineskip}
\noindent\textit{Results}. Rather than seeking an initial solution that
leads to a gravastar at the end of the collapse, it is far easier to
start from a time when the collapse has terminated and the gravastar has
just been formed, and integrate back Eq.~\eqref{dtau-dT} to find the set
of conditions that lead to a successful configuration. Indeed, because
tuning is required to ensure that the de Sitter surface comes to a rest
when the gravastar is eventually produced, integrating back in time is
essential to guarantee that the desired behavior is achieved.

Hence, our ``initial'' conditions at the ``final'' time $T_{\star}$ and
$\eta_{\star}$ can be expressed in terms of the proper circumferential
radii $\mathcal{R}_{1,2}$, \ie
\begin{align}
  \mathcal{R}_{1,\star}&\defeq a_{\rm I}(T_{\star})R_1 = 2M \,,
  \label{initial-condition-1}\\
  \mathcal{R}_{2,\star}& \defeq a_{\rm II}(\eta_{\star})\rho_2=2M +
  \epsilon\,,
  \label{initial-condition-2}
\end{align}
where $\epsilon \lll 1$ in Eq.~\eqref{initial-condition-2} is introduced
because we want the outer shell to be just outside the Schwarzschild
radius, but also to avoid numerical errors when approaching the
divergence in Eq.~\eqref{dtau-dT}. Two additional free parameters need to
be fixed for the integration of Eq.~\eqref{dtau-dT} and are given by the
energy density and spatial curvature in region I, \ie $e_{\rm I}$ and
$k_{\rm I}$, respectively \footnote{We can always set $R_1=1$ since it is
rescaled by the scale factor $a_{\rm I}$, set by our choice of $k_{\rm
  I}$.}. Furthermore, Eqs.~\eqref{initial-condition-1} and
\eqref{initial-condition-2} can be rearranged to specify the values of
$T_{\star}$ and $\eta_{\star}$ (see End Matter for explicit
expressions). After a choice for $e_{\rm I}$ and $k_{\rm I}$ is made, and
requiring that $\rho_1 = \rho_{\star} = \rho_2$, we have all the initial
conditions for the backward integration of Eq.~\eqref{dtau-dT}, which we
perform using an implicit Runge-Kutta scheme of fifth order.

\begin{figure}
    \includegraphics{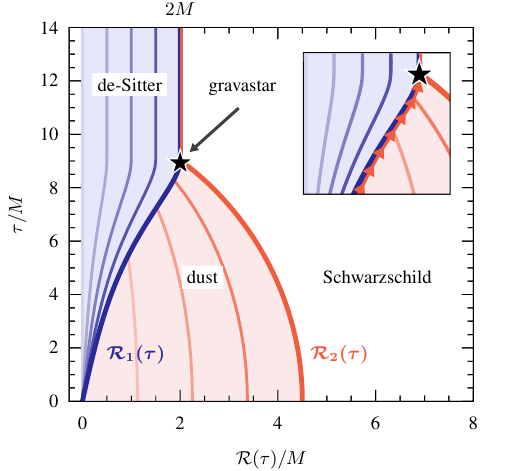}
    \caption{Spacetime reporting the worldlines in proper time of the
      proper circumferential radii $\mathcal{R}_1$ (blue solid line) and
      $\mathcal{R}_2$ (red solid line) separating either the expanding
      de Sitter bubble (blue-shaded area) with the collapsing dust
      (red-shaded area), or the latter with the exterior Schwarzschild
      spacetime. Lines of the same color report worldlines at fractional
      values $1/4$, $1/2$, and $3/4$ of $\mathcal{R}_{1,2}$,
      respectively. Note the formation of a gravastar (black star symbol)
      when $\mathcal{R}_1 = \mathcal{R}_2$. The inset shows the dragging
      of the dust on the de Sitter surface due to the surface tension.}
  \label{fig:fig1}
\end{figure}

Figure~\ref{fig:fig1} offers, via a spacetime diagram, a representative
example of the dynamical formation of a gravastar for $\bar{a}_{\rm
  II}=1$, $\rho_2=(9/2)M$, $\vert k_{\rm I} \vert/k_{\rm II}\simeq 2/3$,
$e_{\rm I} / e_{\rm II}(\eta_{\star}) \simeq 1/3$, and $\epsilon \simeq
10^{-7}$. In particular, shown with a thick solid blue (red) line is the
worldline of the circumferential radius $\mathcal{R}_1 = a_{\rm
  I}(T(\tau))R_1$ [$\mathcal{R}_2 = a_{\rm II}(\tau)\rho_2$] and with
thinner lines the hypersurfaces at fractional values $1/4$, $1/2$, and
$3/4$ of $\mathcal{R}_1$ ($\mathcal{R}_2$). Shown instead with a blue
(red) area is the interior of the expanding de Sitter (collapsing dust)
spacetime. Note that the expanding de Sitter bubble starts from a zero-size
but nonzero initial velocity at $\tau=0$ and that the two worldlines
$\mathcal{R}_1$ and $\mathcal{R}_2$ reach $\mathcal{R}_{1,\star}$ and
$\mathcal{R}_{2,\star}$ at $\tau=\tau_{\star} \defeq \tau(\eta_{\star})$,
with $\mathcal{R}_1$ coming to a smooth stop (\ie $d\mathcal{R}_1 / d\tau
\to 0$). Obviously, the inner part of the collapsing matter will interact
with the edge of the expanding de Sitter bubble leading to a local
increase in density, but not to a shock because of the collisionless
nature of the dust. Furthermore, the presence of a surface tension at the
de Sitter edge will prevent the ``absorption'' of the dust matter, which
is pushed out by the expanding bubble (red arrows in the inset).

\begin{figure}
  \includegraphics{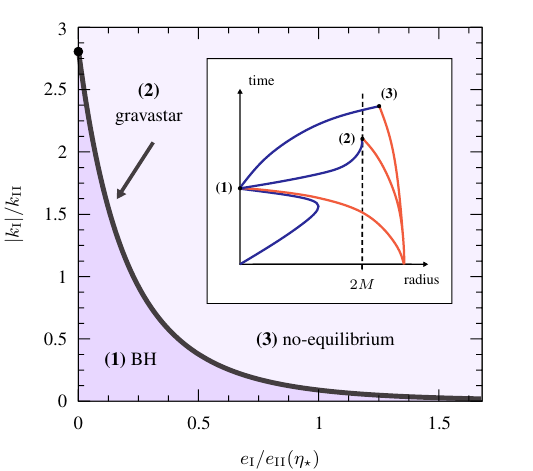}
  \caption{Possible space of parameters for the collapse scenarios when
    expressed in terms of the energy density and spatial curvature of the
    de Sitter region. Shown with a black solid line are the set of
    $(e_{\rm I}, |k_{\rm I}|)$ values that lead to the successful
    creation of a gravastar [case (2)], so that configurations below the
    line lead to either the formation of a black hole [case (1)] or of a
    nonequilibrium configuration [case (3)]. The inset offers a
    schematic spacetime view of the three possible scenarios.}
	\label{fig:fig2}
\end{figure}

Producing a gravastar is only one of the three possible scenarios
resulting from generic initial data [case (2) hereafter]. The other two
possibilities have a final state with $\mathcal{R}_{1,\star} =
\mathcal{R}_{2,\star} = 0$ [case (1) hereafter], or
$\mathcal{R}_{1,\star} = \mathcal{R}_{2,\star} > 2M$ [case (3)
  hereafter]. These cases are summarized in Fig.~\ref{fig:fig2}, which
reports with a solid black line the set of $(e_{\rm I}, |k_{\rm I}|)$
values leading to a successful gravastar formation. Values below the line
(dark-shaded area) refer instead to a black hole formation, while values
above the line (light-shaded area) refer to a configuration that may
correspond (or not) to a stable but less-compact gravastar or to a
nonequilibrium configuration.  Also shown schematically in the inset are
the three possible cases that we have illustrated above, where the
initial position of the de Sitter bubble is chosen only for illustrative
purposes.

\begin{figure*}
	\includegraphics{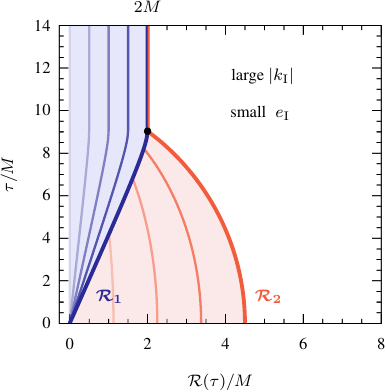}
	\includegraphics{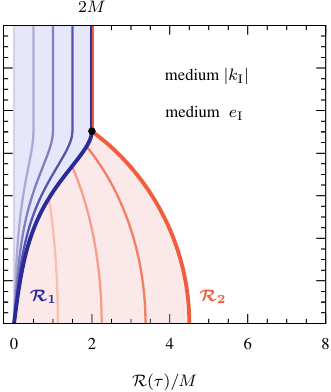}
	\includegraphics{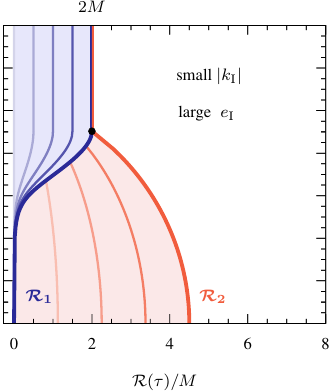}
	\caption{The same as in Fig.~\ref{fig:fig1} but for three sets of
		$(e_{\rm I}, |k_{\rm I}|)$ values referring either to a slow but
		constant de Sitter expansion [large $|k_{\rm I}|$, small $e_{\rm
			I}\to0$; left panel], to an accelerated expansion [medium $|k_{\rm
			I}|$ and $e_{\rm I}/e_{\rm II}(\eta_{\star})\simeq 2/3$; middle
		panel], or to an initially quiescent de Sitter bubble that
		experiences a late and very rapid expansion [small $|k_{\rm I}|$,
		large $e_{\rm I}/e_{\rm II}(\eta_{\star})\simeq 10$; right panel].}
	\label{fig:fig3}
\end{figure*}

Clearly, Fig.~\ref{fig:fig2} shows that a gravastar is produced only for
initial values of the energy density and spatial curvature lying on the
separatrix between the different cases. In this sense, a point on the
separatrix marks the only two values of $e_{\rm I}$ and $|k_{\rm I}|$ for
which a gravastar can be produced, so that the separatrix cannot be an
attractor of the solution. Rather, every point on the separatrix
represents an ``infinitely tuned'' set of initial conditions leading to
the formation of a gravastar. At the same time, because that point is
also part of a one-parameter family of possible initial conditions, there
are infinite sets of initial conditions leading to a gravastar. Stated
differently, while the formation of a single gravastar is infinitely
tuned, there are infinite different ways in which a gravastar can be
produced. The properties of the space of parameters in
Fig.~\ref{fig:fig2} are still largely unknown and future work could
explore whether a critical behavior is exhibited by solutions approaching
the separatrix between cases (1) and (3), along which a gravastar
formation is possible.

A series of remarks is worth making about Fig.~\ref{fig:fig2}. First,
while the considerations made apply in general, the specific values
reported refer to a specific initial radius of the dust sphere, \ie
$\bar{\mathcal{R}}_{2} \defeq \bar{a}_{\rm II} \rho_2 = (9/2)M$. Second,
in general relativity case (1) necessarily leads to a black hole
formation, but it can lead to a bouncing solution covered by an outer
event horizon in alternative theories of gravity. Third, the energy
density and the spatial curvature play a complementary role in the
formation of the gravastar. As we discuss below, this has a direct impact
on the speed at which the de Sitter bubble expands. Finally, while
$|k_{\rm I}| \to 0$ when $e_{\rm I} \to \infty$, the opposite limit of
$e_{\rm I} \to 0$ is particularly interesting and does not correspond to
a divergent value of $k_{\rm I}$. Rather, it represents a Milne
solution~\cite{Milne1935}, that is, a flat Minkowski spacetime in
hyperspherical coordinates or, alternatively, a FLRW solution with zero
energy density, pressure, and cosmological constant (filled circle in
Fig.~\ref{fig:fig2}).

To illustrate the complementary role played by the initial spatial
curvature and energy density of the de Sitter region, we present in
Fig.~\ref{fig:fig3} three representative cases sharing the same
$\bar{\mathcal{R}}_{2}$ but where we vary the values of $\vert k_{\rm
  I}\vert$ and $e_{\rm I}$. More specifically, the left panel refers to a
situation with large $\vert k_{\rm I}\vert$ and small $e_{\rm I}\to0$---note
how the de Sitter edge expands with a nonzero but small initial
velocity, which is kept constant for most of the evolution until it slows
down just before meeting the surface of the collapsing dust sphere. The
middle panel, instead, shows an evolution for intermediate values of
$\vert k_{\rm I}\vert$ and $e_{\rm I}/e_{\rm II}(\eta_{\star}) \simeq
2/3$, that is very similar to that in Fig.~\ref{fig:fig1} and for which
we note that in this case the de Sitter edge has a smaller initial
velocity but also that its evolution is accelerated in the later
stages. Finally, the right panel of Fig.~\ref{fig:fig3} shows the
evolution for small $\vert k_{\rm I}\vert$ and large $e_{\rm I}/e_{\rm
  II}(\eta_{\star})\simeq 10$, and exhibits a rather intriguing behavior
whereby the de Sitter bubble does not expand but at very late times and
it does so in a very rapid manner. This suggests that for $\vert k_{\rm
  I}\vert \to 0$ and $e_{\rm I} \to \infty$ it is possible to construct a
gravastar where the collapse of matter takes place in a fully ordinary
manner until when the stellar surface is very close to $\mathcal{R}_2
\gtrsim 2M$, at which point a de Sitter bubble is nucleated and expands
extremely rapidly, meeting the dust surface at $\mathcal{R}_2 = 2M$.

\vspace{\baselineskip}
\noindent\textit{Compactness Limit}. A well-known result from the
OS collapse is that the proper timescale for the formation of a black
hole of mass $M$ from a dust sphere of initial radius
$\bar{\mathcal{R}}_{2}$ is $\tau = (\pi/2) [\bar{\mathcal{R}}^3_{2} /
  (2M)]^{1/2}$. For any compact object with $\bar{\mathcal{R}}_{2} > 2
M$, $\tau > \pi M$, so that dust clouds with compactness $\mathcal{C}:= M
/ \bar{\mathcal{R}}_{2} \lesssim 1/2$ are allowed. This logic no longer
applies when producing a gravastar, since decreasing the initial dust
radius forces the de Sitter expansion to occur faster, which can be
accomplished by increasing $|k_{\rm I}|$, but only up to a limit. Stated
differently, a causal threshold restricts the compactness of the initial
dust sphere. To see this, we note that a photon emitted at $\eta=0$ at
the center of the dust sphere will take a time
$\eta_{\gamma}=\sqrt{1/k_{\rm II}}\,\, {\rm{arcsin}} \left(\sqrt{k_{\rm
    II}}\rho_2\right)$ to reach the surface of the dust sphere at
$\rho=\rho_2$. The compactness bound then follows from determining the
initial dust radius that will be reached by such a photon before the dust
surface reaches the Schwarzschild radius, \ie $\eta_{\gamma} \leq
\eta_{\star}$. This constraint sets the maximal compactness to be
\begin{equation}
  \mathcal{C} = \frac{M}{\ShiftDown{0.2em}{$\bar{\mathcal{R}}_2$}} \leq
  \frac{3}{8} = 0.375\,,
\end{equation}
which is only slightly smaller than the Buchdahl limit of
$\mathcal{C}_{_{\rm B}} = 4/9 \simeq 0.444$~\cite{Buchdahl:59}.
Interestingly, the limit $\mathcal{C} \to 3/8$ also requires that
$|k_{\rm I}| \to \infty$ in order for the de Sitter bubble to be able to
stop the collapse.\pagebreak

\noindent\textit{Conclusion}. We have here provided a simple but
comprehensive answer to a long-standing question: how is a gravastar
formed? Starting from the OS collapse, we have embedded an expanding
de Sitter bubble within the collapsing dust sphere. As a result, the
evolving spacetime is split into three different regions patched together
by junction conditions guaranteeing the continuity of the induced
metric. In this way, it is possible to combine the relevant Friedmann
equations for the two regions into a master differential equation that
describes the three families of solutions reflecting the possible
outcomes of the collapse process. In particular, we have shown that,
depending on the initial conditions, the de Sitter expansion and dust
collapse can either lead to a black hole, to a nonequilibrium
configuration, or to a static gravastar. This latter outcome requires a
fine-tuning of the initial spatial curvature and energy density in the
de Sitter region, but there is an infinite family of initial conditions
specified by a set of $(e_{\rm I}, | k_{\rm I} |)$ leading to the same
static gravastar, although with different dynamics.

More specifically, large values of $\vert k_{\rm I}\vert$ and small
values of $e_{\rm I}$ lead to a de Sitter bubble that expands with a
nonzero but small initial velocity, which remains constant for most of
the evolution before slowing down at gravastar formation. By contrast,
small $\vert k_{\rm I}\vert$ and large $e_{\rm I}$ lead to a de Sitter
bubble that expands only at very late times and it does so in a very
rapid manner. This latter case suggests an appealing scenario in which
the collapse of matter takes place in a fully ordinary manner until the
stellar surface is very close to $2M$, at which point the de Sitter
bubble can nucleate---\eg as a result of quantum fluctuations, conformal
anomaly~\cite{Mottola06, Mottola2023b, Mottola2025}, or instabilities in
an anti-de Sitter spacetime~\cite{Biasi2022}---and very rapidly stops
the collapsing dust surface, forming a gravastar. Interestingly, because
the whole process of gravastar formation cannot be superluminal, a
maximum compactness of $\mathcal{C} \leq 3/8$ emerges for the dust
sphere, above which the collapse cannot produce a gravastar but a black
hole.

The dynamics presented here offers, for the first time, a rather
straightforward example of how, remaining within general relativity and
without invoking higher-curvature corrections, it is possible to avoid
both singularity and black hole formation by employing the repulsive role
of a de Sitter solution. At the same time, while our work shows that
gravastar formation is possible, future studies will have to determine
whether this scenario will continue to be possible under more realistic
assumptions, such as the use of more realistic equations of state for the
collapsing matter, the nucleation of de Sitter bubbles away from the
center (potentially leading to a ``nestar''
solution~\cite{Jampolski2024}), or deviations from spherical symmetry
that would lead to potential instabilities of the
shell~\cite{Yang2023b}. More importantly, additional studies will have to
elucidate whether generic initial conditions in a gravitational collapse
are more likely to lead to a gravastar or to a black hole.

\vspace{\baselineskip}
\noindent\textit{Acknowledgements}. It is a pleasure to thank A. Bonanno,
F. Camilloni, F. Di Filippo, E. Mottola, and A. Murcia-Gil for useful
discussions. Partial funding comes from the ERC Advanced Grant
``JETSET: Launching, propagation and emission of relativistic jets
from binary mergers and across mass scales'' (Grant No. 884631).
D.J. acknowledges support by the Hans-B\"ockler-Stiftung (German Academic
Scholarship Foundation).  L.R. acknowledges the Walter Greiner Gesellschaft
zur F\"orderung der physikalischen Grundlagenforschung e.V. through the
Carl W. Fueck Laureatus Chair.

\vspace{\baselineskip}
\noindent\textit{Data Availability}. The data that supports the findings of
this article are not publicly available. The Data are available from the
authors upon reasonable request.

\begin{figure}
	\includegraphics{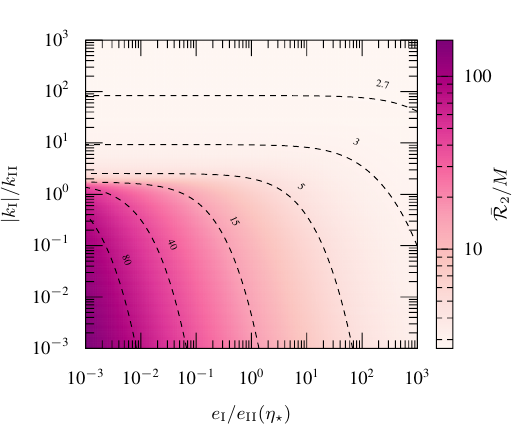}
	\caption{The set of parameters $(e_{\rm I},| k_{\rm I} |)$ leading to
		[case (2)] a gravastar formation, with different colors representing
		different initial dust radii $\bar{\mathcal{R}}_2$. Shown with dashed
		lines are contours corresponding to initial radii with
		$\bar{\mathcal{R}}_2/M=[2.7,3,5,15,40,80]$.}
	\label{fig:emfig1}
\end{figure}

\vspace{\baselineskip}
\noindent\textit{Generalization to arbitrary dust configurations}.
While we have so far considered the formation of a gravastar from a
specific dust configuration with $\bar{\mathcal{R}}_2/M = 9/2$, it is
indeed possible to form a gravastar from an arbitrary initial radius so
long as $\mathcal{C} \leq 3/8$. The parameter space in
Fig.~\ref{fig:fig2} can thus be extended to arbitrary initial radii
$\bar{\mathcal{R}}_2/M \geq 8/3$, whilst only looking at case
(2). Figure~\ref{fig:emfig1} reports with a color map the set of
parameters $(e_{\rm I},| k_{\rm I}|)$ leading to the formation of a
gravastar, with different colors representing different initial radii
$\bar{\mathcal{R}}_2$. Shown with dashed lines are the contours
corresponding to initial radii with $\bar{\mathcal{R}}_2/M = [2.7,3,5,15,
  40, 80]$. Note that increasing $e_{\rm I}$ only slowly decreases the
initial radii of the corresponding dust sphere, while a steep decrease
happens when varying $k_{\rm I}$ for $|k_{\rm I} | / k_{\rm II}\in(1,2)$.

\begin{figure}
  \hspace{-1.5em}\includegraphics{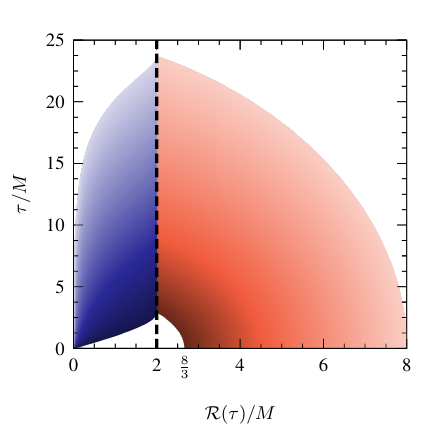}
  \caption{Worldlines $\mathcal{R}_1(\tau)$ and
    $\mathcal{R}_2(\tau)$ for initial radii
    $\bar{\mathcal{R}}_2/M\in(8/3,8)$ being represented by
    different colors for a fixed de Sitter energy density $e_{\rm
      I}/e_{\rm II}(\eta_{\star})=1$, which in turn determines the
    curvature constant $k_{\rm I}$.}
  \label{fig:emfig2}
\end{figure}

\begin{figure*}
  \hspace{-1.0em}\includegraphics{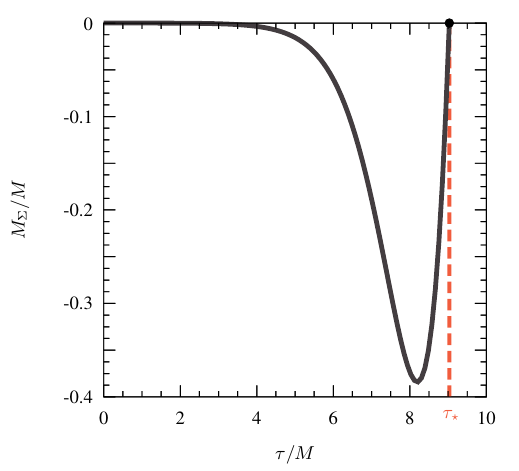}
  \includegraphics{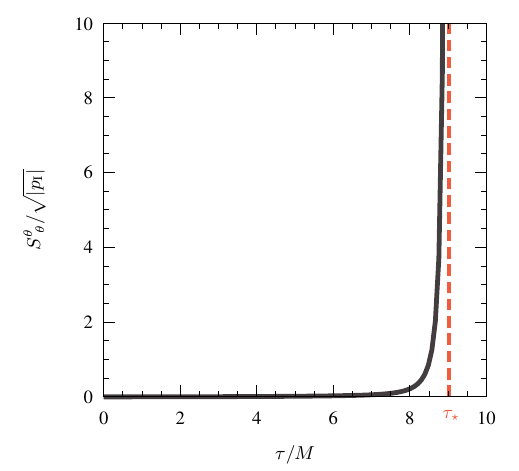}
  \caption{Left panel: evolution of the surface mass
    $M_{\Sigma}$ for the same dust parameters as in Fig.~\ref{fig:fig1}
    but with a de Sitter energy density $e_{\rm I}/e_{\rm
      II}(\eta_{\star})=1$, leading to a vanishing surface energy density
    at the end of the collapse at $\tau_{\star}$. Right panel:
    evolution of the surface tension $S^{\theta}_{~\theta}$, which builds
    up from zero, when the de Sitter expansion has not yet started, till
    divergence, when the dust gets compressed to zero thickness, as in
    a static gravastar.}
  \label{fig:emfig3}
\end{figure*}

Figure~\ref{fig:emfig2} depicts with different colors the collapse to a
gravastar for initial radii $\bar{\mathcal{R}}_2/M\in(8/3,8)$ and the
corresponding motion of the hypersurfaces $\mathcal{R}_1(\tau)$ and
$\mathcal{R}_2(\tau)$ for a fixed de Sitter energy density $e_{\rm
  I}/e_{\rm II}(\eta_{\star})=1$, which in turn determines the curvature
constant $k_{\rm I}$. As mentioned in the main text, the timescale for
the formation of a gravastar can be computed via
\eqref{initial-condition-1}--\eqref{initial-condition-2}, leading to
\begin{align}
  T_{\star} &= \sqrt{{3}/{8\pi e_{\rm I}}}\,{\rm arsinh}
  \left({2M\sqrt{{8\pi{e}_{\rm I}}}}/\sqrt{{3|k_{\rm
        I}|}}{R_1}\right)\,,\\
  \eta_{\star} &= \sqrt{1/k_{\rm II}}\,{\rm arccos}
  \left([4M+2\epsilon]/\bar{\mathcal{R}}_2-1\right)\,,
\end{align}
which can be re-expressed in terms of $\tau_{\star}$.

\vspace{\baselineskip}
\noindent\textit{Surface energy density and tension}. Within the
thin-shell description of two adjacent spacetimes, it is possible to
define the surface stress-energy tensor
\begin{equation}
8 \pi S_{\mu\nu} = A_{\mu\nu} - \frac{1}{2} A g_{\mu\nu}\,,
\end{equation}
where $A_{\mu\nu}$ is the four-dimensional extrinsic curvature and $A
\defeq A^{\mu}_{~\mu}$ is its trace (see, \eg \cite{Israel1966, Poisson04a,
  Rezzolla_book:2013} for details). The discontinuity of the extrinsic
curvature gives rise to a surface stress-energy three-tensor $S_{ij}$
defined only on the de Sitter hypersurface
\begin{equation}
  \label{eq:Sij}
  8 \pi S_{ij} = \mp \left( [K_{ij}] - [K]h_{ij} \right)\,,
\end{equation}
where $h_{ij}$ is the induced metric, $K_{ij}$ is the extrinsic curvature,
$K \defeq K^i_{~i}$ is its trace, and the square brackets measure a jump
across the values $\psi^{\pm}$ of any quantity on either side of the
hypersurface, \ie $[\psi] \defeq \psi^+ - \psi^-$. In our convention,
``$-$'' denotes region I (de Sitter) while ``$+$'' region II (dust). We
should also remark that the surface stress-energy three-tensor $S_{ij}$
can be seen as measuring the jump in the extrinsic curvature across the
surface $\Sigma$ separating the de Sitter and the dust region---hence as
a purely geometrical quantity---or as measuring the jump in the
stress-energy four-tensor across the same surface. In this latter case,
it does not have a well-defined underlying microphysical description and
hence a clear equation of state. It is therefore debatable whether such a
virtual form of matter should obey an energy condition and whether a
violation of an energy condition has any physical implication.

Choosing $(T,\theta,\phi)$ as the coordinates on the de Sitter
hypersurface, the nonzero components of the induced metric are given by
\begin{align}
  h_{TT}^{-}&=-1\,,\\ h_{\theta\theta}^{-}&=a_{\rm I}^2
  R_1^2\,,\\ h_{TT}^{+}&=-\dot{\tau}^2+a_{\rm II}^2\dot{\rho}_1^2/x
  \,,\\ h_{\theta\theta}^{+}&=a_{\rm II}^2\rho_1^2\,,
\end{align}
where $x$ is an auxiliary variable defined as $x\defeq 1-k_{\rm
  II}\rho_1^2$, while the nonzero components of the extrinsic curvature
tensor are given by
\begin{align}
  K_{\theta\theta}^{-}&=a_{\rm I} R_1 \sqrt{1+|k_{\rm
      I}|R_1^2}\,,\\ K_{TT}^{+}&=-x^{-3/2} \left[k_{\rm II}a_{\rm
      II}\rho_1\dot{\rho}_1^2\dot{\tau} -a_{\rm
      II}^2\partial_{\tau}a_{\rm II}\dot{\rho}_1^3 \right.\notag\\ &
    \phantom{=} \left. -x\left(a_{\rm II}\dot{\rho}_1\ddot{\tau} -a_{\rm
      II}\ddot{\rho}_1\dot{\tau} -2\partial_{\tau}a_{\rm
      II}\dot{\rho}_1\dot{\tau}^2
    \right)\right]\,,\\ K_{\theta\theta}^{+}&=a_{\rm II}\rho_1 x^{-1/2}
  \left(x\dot{\tau} +a_{\rm II}\partial_{\tau}a_{\rm
    II}\rho_1\dot{\rho}_1 \right)\,,
\end{align}
where a dot denotes a derivative with respect to $T$, so that $\dot{\tau}
= \dot{\eta}a_{\rm II}$ and $\ddot{\tau} = \ddot{\eta}a_{\rm
  II}+\dot{\eta}^2 a^{\prime}_{\rm II}$. Finally, due to spherical
symmetry, $h_{\phi\phi}^{\pm}=h_{\theta\theta}^{\pm}\,{\rm sin}^2\theta$
and $K_{\phi\phi}^{\pm}=K_{\theta\theta}^{\pm}\,{\rm sin}^2\theta$.

The $S^T_{~T}$ and $S^{\theta}_{~\theta}$ components of the surface
stress-energy tensor correspond, respectively, to the surface energy
density and surface tension on the de Sitter hypersurface. While the
surface tension is unavoidable, as it stabilizes the construction via a
tangential pressure, the surface energy density can be chosen to be zero
at the end of the collapse. We can see this by inspecting the quasilocal
Misner-Sharp mass $dm/d\mathcal{R} = 4\pi\mathcal{R}^2e$, and looking at
the distributions of the different regions, being $M = M_{\rm I} +
M_{\Sigma} + M_{\rm II}$, with $M_{\rm I} \defeq (4\pi/3) e_{\rm
  I}\mathcal{R}^3_1$, and $M_{\rm II} \defeq (4\pi/3) e_{\rm
  II}(\mathcal{R}^3_2 - \mathcal{R}^3_1)$. Since the energy density is
homogeneous in all regions, the Misner-Sharp mass can easily be
evaluated, and rearranged with respect to the surface mass
\begin{equation}
  \label{eq:Msigma}
  M_{\Sigma} = \frac{4\pi}{3}\mathcal{R}_1^3\left(e_{\rm II}-e_{\rm
    I}\right)\,,
\end{equation}
which is proportional to the surface energy density $M_{\Sigma} \propto
S^T_{~T}$, so that if the surface energy density vanishes, the surface
mass does so too.

The evolution of the surface mass $M_{\Sigma}$ is shown in the left panel
of Fig.~\ref{fig:emfig3}, while that of the surface tension
$S^{\theta}_{~\theta}$ is reported in the right panel. The initial
conditions for the dust parameters are the same as in
Fig.~\ref{fig:fig1}, but the initial energy density in the de Sitter
region is set to be $e_{\rm I}/e_{\rm II}(\eta_{\star})=1$, which leads
to negative values during the collapse and to $M_{\Sigma}= 0$ at the end
of the collapse. In this regard, we should note that what is relevant for
an external observer is the total mass $M$ of the system, which is
positive and remains constant. Splitting such a mass in $M_{\rm I}$,
$M_{\rm II}$, and $M_{\Sigma}$---which are derived quantities and do not
have fundamental equations governing their dynamics---is done to recast
these quantities in terms of an intuitive representation (\eg the mass
enclosed in a sphere of a radius $R_{\rm I}$ and $R_{\rm II}$), so that
$M_{\Sigma}$ is just the difference between the total mass $M$ and the sum
of $M_{\rm I}$ and $M_{\rm II}$, and can be negative.  Indeed, $M_\Sigma$
is not restricted to negative values and different initial conditions can
lead to positive values throughout the gravastar formation process.

The fact that the surface mass becomes negative during the evolution can
be explained in a number of different ways. First, to keep the total mass
$M$ constant while the mass of the de Sitter region $M_{\rm I}$ increases
more rapidly than the decrease of the mass of the dust region $M_{\rm
  II}$. Alternatively, as a result of the choice of the initial (positive)
energy density, which is $e_{\rm I} > e_{\rm II}$ essentially at all
times and provides a negative contribution in Eq.~\eqref{eq:Msigma}. The
behavior of the surface tension is instead simpler to understand and
increases from an initial zero value, when the expansion of the de Sitter
bubble has not started yet, diverging when the dust shell is compressed
to zero thickness. This is in perfect agreement with what is expected for a
static gravastar, where $S^{\theta}_{~\theta} \to \infty$ in the limit of
a zero-thickness shell, \ie for $\epsilon\to 0$.

\end{document}